\documentclass[aip,reprint,onecolumn,english,nofootinbib,letterpaper,nobalancelastpage]{revtex4}
\usepackage{lmodern}

\usepackage[T1]{fontenc}
\usepackage[latin9]{inputenc}
\usepackage[letterpaper]{geometry}
\usepackage{amsmath}
\usepackage{amssymb}
\usepackage{bbm}
\usepackage{babel}
\usepackage{graphicx}

\usepackage{latexsym}
\usepackage{mathptm}

\newcommand{\f}{\begin{equation}}
\newcommand{\ff}{\end{equation}}

\begin{document}

\title{
{ Relative locality: A deepening of the relativity principle}\\
}
\author{$~$\\
{\bf Giovanni Amelino-Camelia}$^{1,a}$,
  {\bf Laurent Freidel}$^{2,b}$, {\bf Jerzy Kowalski-Glikman}$^{3,c}$, {\bf Lee Smolin}$^{2,d}$
\\
$^1${\footnotesize Dipartimento di Fisica, Universit\`a ``La Sapienza" and Sez.~Roma1 INFN, P.le A. Moro 2, 00185 Roma, Italy }\\
$^2${\footnotesize Perimeter Institute for Theoretical Physics, 31 Caroline Street North, Waterloo, Ontario N2J 2Y5, Canada}\\
$^3${\footnotesize Institute for Theoretical Physics, University of Wroclaw,  Pl. Maxa Borna 9, 50-204 Wroclaw, Poland}\\
$~$\\$~$\\$~$\\$~$ }

\begin{abstract}
{\normalsize {\bf SUMMARY}:
We describe a recently introduced principle of relative locality which we
propose governs a regime of quantum gravitational phenomena
accessible to experimental investigation. This regime comprises phenomena in
which $\hbar$ and $G_N$   may be neglected, while their ratio,
the
Planck mass $M_p =\sqrt{\hbar / G_N}$, is important. We propose that $M_p$
 governs the scale at which momentum space may have a curved geometry. We find that
there are striking consequences for the concept of locality.  The
description of events in spacetime now depends on the energy used to probe
it.  But there remains an invariant description of physics in phase space.
There is furthermore a reasonable expectation that the geometry of momentum
space can be measured experimentally using astrophysical observations.
}
 \end{abstract}

\maketitle

\vskip 2.7cm

\begin{center}
{\it This essay was awarded Second Prize in the 2011 Essay Competition of the
Gravity Research Foundation.}
%\vskip 0.5 cm submission date: 31 March 2011
\end{center}

\vskip 4.6cm

\noindent
{\footnotesize{ \indent \indent $^a${\tt amelino@roma1.infn.it}\\
\indent \indent $^b${\tt
lfreidel@perimeterinstitute.ca}\\
\indent \indent $^c${\tt
jkowalskiglikman@ift.uni.wroc.pl}\\
\indent \indent $^d${\tt
lsmolin@perimeterinstitute.ca}}}

\newpage
{\large
\baselineskip 18pt

How do we know we live in a spacetime? And, if so, how do we know we
all share the same spacetime? According to the operational  procedure
introduced by Einstein~\cite{Einstein}, we infer the coordinates of a distant event by
analyzing light signals sent between observer and the event. But when we do this we throw away
information about the energy of the photons.  This is clearly a good approximation, but is it
exact?  Suppose we use Planck energy photons or red photons in Einstein's localization procedure,  can we be
sure that  the spacetimes we infer in the two cases are going to be the same?
Also, how can we be sure that
 when two events are inferred to be at the same
spacetime position by one observer, the same holds true
for another, distant  observer?

In special and general  relativity the answer to these questions is yes. Simultaneity is relative but locality is
absolute. This follows from the assumption that spacetime is a universal entity in which all of physics unfolds.
However, all approaches to the study
of the quantum-gravity problem suggest that locality
must be  weakened and that the concept of spacetime is only emergent
and should be replaced by something more fundamental.
A natural and pressing question is whether it is possible to relax the universal locality assumption in a controlled manner, such that
it gives us a stepping stone toward the full theory of quantum gravity?

A natural  guess  is that
the Planck length\footnote{We work in units such that the speed-of-light scale $c$
is set to 1.}, $\ell_p = \sqrt{\hbar G}$,
sets an absolute limit to how precisely an event can be  localized,
$\Delta x \sim \ell_p.$
However,  the Planck length is non zero only if $G$ and $\hbar$ are non zero, so this hypothesis requires a full fledged quantum gravity theory to elaborate it.
But there is an alternative, which is to explore a ``classical-non gravitational'' regime of quantum gravity which still captures some of the
key delocalising features of quantum gravity. In this regime,
$\hbar$ and $G$ are both neglected, while their ratio
 is held fixed:
\f
\hbar \rightarrow 0~,~~~G_{N} \rightarrow 0~,~~~\mathrm{but~with~fixed}~ \sqrt{\frac{\hbar}{G_{N}}}=M_p
\label{thelimit}
\ff
In this regime of quantum gravity, which is labeled the  ``relative-locality regime" in the recent Ref.~\cite{AmelinoCamelia:2011bm},
both quantum mechanics and gravity are switched off, but we still keep effects due to the presence of the Planck mass.  Remarkably,
as we will describe, this regime includes effects on very large scales which can be explored in astrophysical experiments~\cite{GioLee,next}. Furthermore,
since $\hbar$ and $G_N$ are both zero it can be investigated in simple phenomenological models.

\begin{figure}[h!]
\begin{center}
\includegraphics[width=0.51 \textwidth]{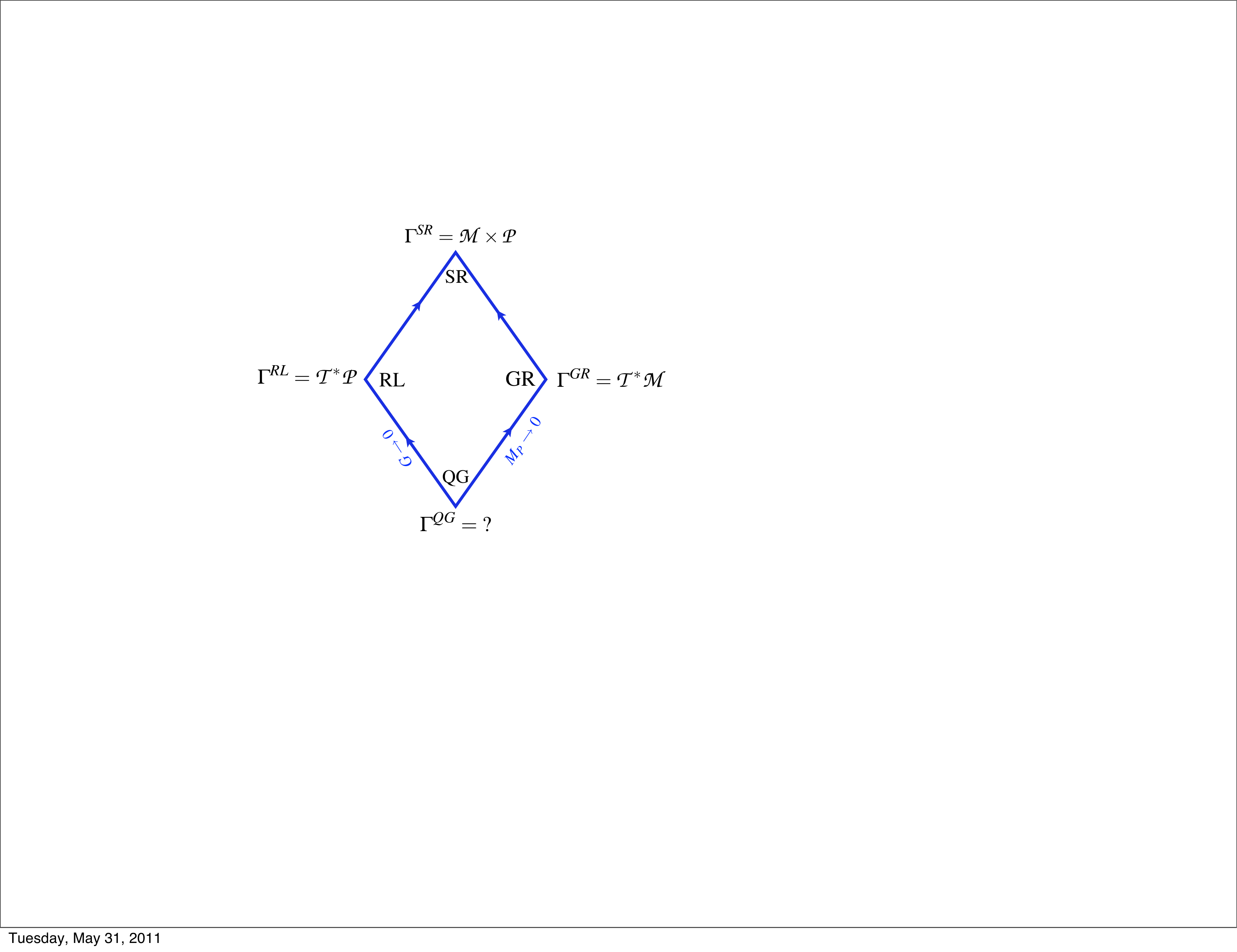}
\caption{We show here that general relativity and relative locality are two ways of deepening the relativity principle.  In general relativity spacetime is curved but momentum space is flat. The opposite is the case in relative locality.  This has consequences for the phase space description as is shown, and elaborated below.
Alternatively, starting from an unknown quantum theory of gravity,  one can ascend to special relativity through two paths.
Taking $\hbar \rightarrow 0$ but keeping
$G_N$ fixed (so that $M_p $ also goes to $ 0$) one ascends on the right to general relativity. But there is an alternative.  Keep $M_p$ fixed while taking $G\rightarrow 0$ (and hence also $\hbar \rightarrow 0$) leads to the relative locality regime on the left. }
\end{center}
\end{figure}

In Ref.~\cite{AmelinoCamelia:2011bm}  we show that the hypothesis of universal locality is equivalent to the
statement that momentum space is a linear space. It is natural then to propose  that the mass scale $M_{P}$ parameterizes non linearities in momentum space.
Remarkably, these non linearities can be understood as introducing on momentum space a {\it non trivial  geometry}. In ~\cite{AmelinoCamelia:2011bm} we introduced
a precise formulation of the geometry of momentum space from which the consequences for the questions we opened with can be exactly derived.

The idea that momentum space should have a non trivial geometry when quantum gravity effects are taken into account was  originally proposed by Max Born,
as early as 1938 \cite{Born1938}.
He argued that the validity of quantum mechanics implies there is in physics an equivalence between
space and momentum space, which we now call  Born reciprocity.
The introduction of gravity breaks this symmetry between space and momentum space because space is now curved while
momentum space is a linear space-and hence flat. Allowing the momentum space geometry to be curved is a natural way to reconcile
gravity with quantum mechanics from this perspective.

Remarkably, this is exactly what has been shown to happen in a very illuminating
toy model of quantum gravity, which is quantum gravity in $2+1$ dimensions coupled to matter.  There
 Newton's constant $G$ has dimensions of inverse mass, and indeed it
turns out~\cite{Matschull:1997du,Freidel:2005me}
that in $2+1$  dimensions the momentum
space of particles and fields is a manifold of constant curvature
$G^2$, while spacetime is (locally) flat~\cite{deserjackiw}.

There are two kinds of non-trivial geometry (metric and connection) any manifold, including
momentum space,  can have. Each of these has, as shown in~\cite{AmelinoCamelia:2011bm},
 a characterization in terms of observable properties for  the dynamics of particles.
A  metric in momentum space $ds^{2}= g^{\mu\nu}(p) {\mathrm d}p_{\mu}{\mathrm d}p_{\nu}$ is needed  in order to write
energy-momentum on-shell relation
\f
m^{2} = D^{2}(p)
\ff
where $D(p)$ is the distance of the point $p_{\mu}$ from the origin $p_{\mu}=0$.
A non-trivial affine connection  is needed in order to  produce
 non-linearities in the law of composition of momenta, which is
used in formulating the conservation of momentum
\f
\left(p \oplus q\right)_\mu \simeq p_\mu + q_\mu -
\frac{1}{M_p} \Gamma_\mu^{\alpha\beta}\, p_\alpha\,
q_\beta + \cdots
\ff
where on the right-hand side we assumed momenta are small with respect
to the Planck mass $M_p$ and $\Gamma_{\mu}^{\alpha\beta}$
 are the ($M_p$-rescaled) connection coefficients on momentum space
 evaluated at $p_\mu = 0$.

We can show that the geometry of momentum space has a profound effect on localisation  through an elementary argument.  To do this
 we look at the role that
the special-relativistic linear law of conservation of momenta
has in ensuring that locality is absolute.
Suppose $x^\mu_I$ are the positions of several particles that coincide at
the event $e$ in the coordinates of a given observer.
The total-momentum conservation law generates the transformation
from that observer to another separated
 from the first observer by a vector, $b^\mu$.
  In the special relativistic case the total momentum is the linear
  sum $ P_\nu^{tot} = \sum_J p_\nu^J$ and one finds

\f
\delta x^\mu_I = \{ x_I^\mu , b^\nu { P }^{tot}_\nu \}
= \{ x_I^\mu , b^\nu \sum_J p^J_\nu \}
= b^\mu
\ff
so that all the worldlines are translated together,
independent of the momentum they carry.

This is the familiar notion of absolute locality afforded
by the special-relativistic setting.
If instead momentum space has a non-trivial connection,
in the sense discussed above,
then
%${\cal P }^{total}_\mu$ is nonlinear, {\it i.e.},
${ P }^{total}_\mu$ is nonlinear, {\it i.e.},
\f
{ P }^{total}_\mu =\sum_I p^I_\mu
+ \frac{1}{M_p} \sum_{I<J} \Gamma_{\mu}^{\nu \rho} p_\nu^I p_\rho^J
\ff
 Then
\f
\delta x^\mu_I = \{ x_I^\mu , b^\nu {P }^{\mathrm{total}}_\nu \}
= b^\mu + \frac{1}{M_p} b^\nu  \sum_{J>I} \Gamma_{\nu}^{\mu \rho}  p_\rho^J.
\ff
Thus we see that how much a worldline of a particle is translated depends on the momenta carried by it and the particles it interacts with.
The net result is the feature we call ``relative locality", illustrated in
Fig.~2.
Processes are still described as local in the coordinatizations of spacetime
by observers close to them, but those same processes are described as nonlocal
in the coordinates adopted by distant observers.

\begin{figure}[h!]
\begin{center}
\includegraphics[width=0.54 \textwidth]{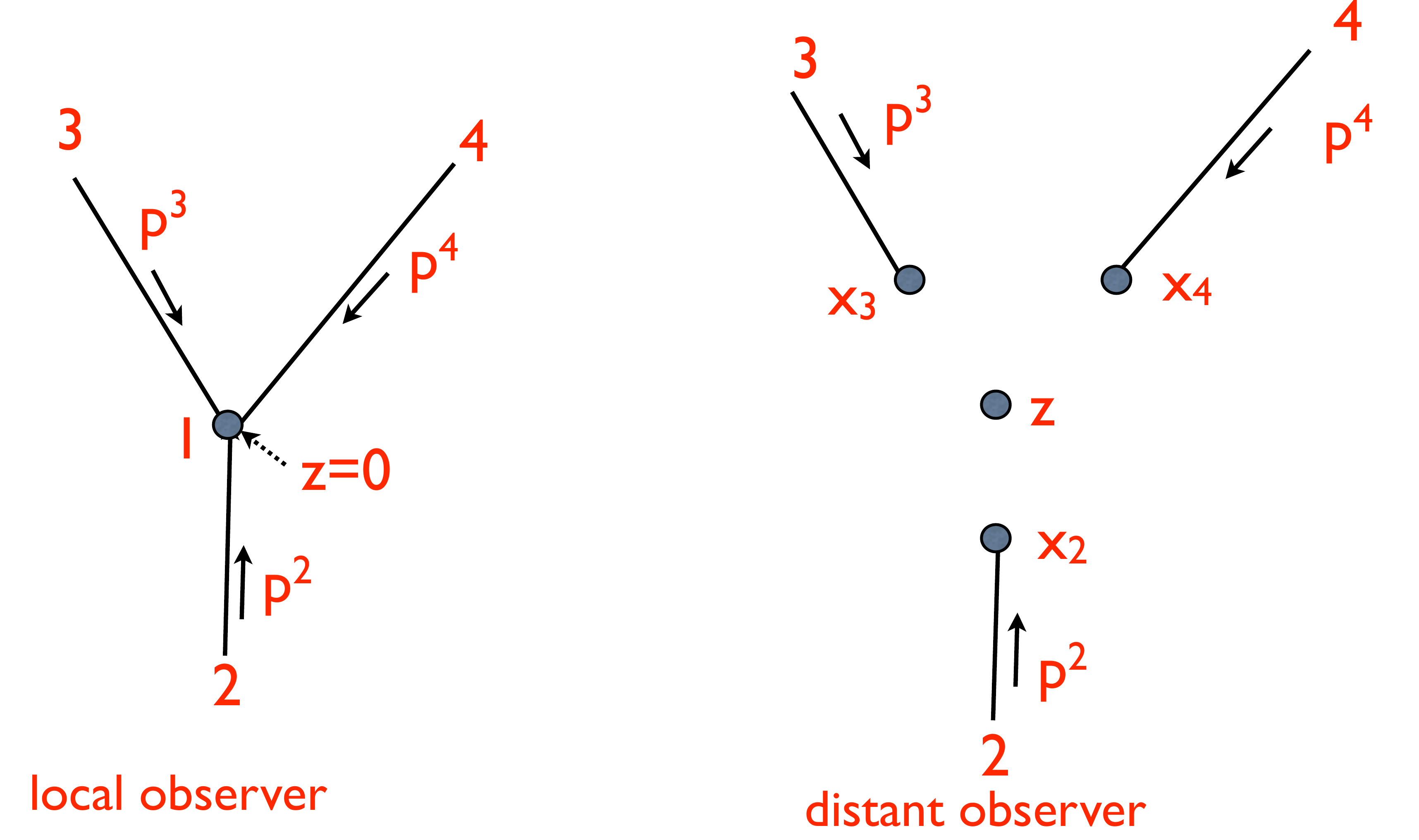}
\caption{{ Relative locality implies that the projection from the invariant phase space description to a description of events in spacetime leaves a picture of
localization which is dependent on the relation of the observer (or origin of coordinates) to the event.  If the event is at the origin of the observer's coordinate system, then the event is described as local, as on the left. But if the event is far from the origin of the observer's coordinates, the event is described as non-local, in the sense that the projections of the ends of the worldlines no longer meet at the point where the interaction takes place.  This is not a weakening of the requirement that physics is local, it is instead a consequence of the energy dependence of the procedure by which the coordinates of distant events are inferred.  There is an invariant description, but it is
in a phase space.}}
\end{center}
\end{figure}

These novel phenomena have a consistent mathematical description in which the notion of spacetime gives way to an invariant geometry formulated in a phase space.   In special relativity, the phase space associated with each particle is a product of spacetime and
momentum space, {\it i.e.} $\Gamma^{SR} = {\cal M} \times {\cal P}$.

In general relativity, the spacetime manifold $\cal M$ has a curved geometry. The particle  phase space is no longer a product.  Instead, there is a separate momentum space,
${\cal P}_x$ associated to each  spacetime point $x \in {\cal M}$. This is identified
with the cotangent space of $\cal M$ at $x$, so
that  ${\cal P}_x = {\cal T}^{*}_{x} ({\cal M})$.  The whole phase space is the cotangent bundle
of $\cal M$, {\it i.e.} $\Gamma^{GR} = {\cal T}^{*}({\cal M} )$

Within the framework of relative locality, it is the momentum space $\cal P$ that is curved.  There then must be a separate spacetime, ${\cal M}_p$ for each value
of momentum, ${\cal M}_p = {\cal T}^{*}_{p}({\cal P } )$.  The whole phase space is then the cotangent bundle over momentum space, {\it i.e.} $\Gamma^{RL} = {\cal T}^{*} ({\cal P})$.

If one wants to compare momenta of particles at different points of spacetime in general relativity, $x$ and $y$, one needs to parallel transport the covector
$p_a (x)$ along some path $\gamma$ from $x$ to $y$, using the spacetime connection.
Now, suppose, within the dual framework of relative locality, we want to know if the worldlines of  two particles, $A$ and $B$, with different momenta, meet.
We cannot assert that $x^\mu_A = x^\mu_B$ because, quite literally, they live in different spaces, as they correspond to particles of different momenta. What we can do is to ask that there is a parallel transport on momentum space that takes them to each other.  If so, there will be a linear transformation, $[U_\gamma ]_\nu^\mu $,
which maps the spacetime coordinates associated with momenta $p^A_\mu$
to those associated with the momenta $p^B_\mu$.
This will be defined by the parallel transport along a path $\gamma$
in momentum space, so that
\f
x_B^\mu = [U_\gamma ]_\nu^\mu x_A^\nu
\label{utrasf}
\ff
This can be implemented very precisely from an action principle  associated with every interaction process.
The free part of the action associated with each worldline given by
\f
S^{\mathrm{free}} = \int {\mathrm d} s( x^{\mu} \dot{p}_{\mu} + N ( D^{2}(p)-m^{2}) )
\ff
 imposes the on-shell relation, while the interaction implement the conservation law ${\cal K}(p_{I}(0))=0$ at the interaction event
 \f
 S^{\mathrm{int}} = z^{\mu} {\cal K}_{\,\,\mu}(p_{I}(0)).
 \ff
 The relationship (\ref{utrasf}) follows from the variation of this action principle with respect to the momenta at the inreraction events.
It turns out that the path $\gamma$, along which we parallel transport a spacetime coordinate in momentum space, is specified by the form of the conservation law at an interaction event
between the two particles.  This is very parsimonious, it says that the two particles need to interact if we are to assert whether their worldlines cross.

Notice that according to (\ref{utrasf})
one is still assured that if the event is such that, in the
coordinates of a given observer,  $x_A^\mu=0$ then it is also the case that $x_B^\mu=0$.
This is why we assert that there are always observers, local to an interaction,
who see it to be local.
One also sees that if the connection vanishes then $(U_{\gamma})_{\mu}^{\nu}=\delta_{\mu}^{\nu}$ and $x_{A}=x_{B}$ and we recover the usual
picture where interaction are local.

Let us expand the parallel transport in terms of the connection:
\f
[U_\gamma ]_\nu^\mu  = \delta^\nu_\mu + \frac{1}{M_p} \Gamma_\mu^{\nu \rho}p_\rho + ...
\ff
It will follow that the difference $\Delta x^{\mu}$ between $x_A^\mu$ and $x_B^\mu$
is proportional to $x_A^\mu$ and $p_\mu$. It can therefore be said that the deviation of locality is at first order of the form
\f\label{Deltax}
\Delta x \sim x \frac{E}{M_{P}}.
\ff
We see from this formula (\ref{Deltax}), that
the smallness of $M_p^{-1}$ can be compensated by a large distance $x$,
so that over astrophysical distances values of
$\Delta x$ which are consequences of relative-locality effects take macroscopic values~\cite{next}.
A more detailed analysis shows that   there really are observable effects on these scales~\cite{next} which are relevant for current astrophysical
observations of gamma ray bursts,  in which precise measurements of arrival times are used to set bounds on the  locality of distant events~\cite{GioLee,fermi}.
But this is not all.  Other experiments which may measure
or bound~\cite{usinprep}
the geometry of momentum space at order $M_p^{-1}$  include tests of the linearity
of momentum conservation using ultracold atoms~\cite{jurekmich}
and the development of air showers produced by cosmic rays~\cite{dedenko}.

Such phenomena are very different in nature from the predictions of detailed
quantum theories of gravity for the Planck length regime.  It is unlikely we will ever detect a graviton~\cite{leenograv,gacQM100,nograviton}, but it is
reasonable to expect that relative locality can really be distinguished experimentally from absolute locality.
By doing so the geometry of momentum space can be measured.

A 19th-century scientist conversant with Galilean relativity could have asked: do we ``see" space?
Einstein taught us that the answer is negative: there is  a maximum speed and at
best we ``see" spacetime.  We now argue that this too is wrong.  What
we really see in our telescopes and particle detectors are quanta arriving at different
angles with different momenta and energies.
Those observations allows us to infer the existence of a universal and energy-independent
description of physics in a space-time only if momentum space has a trivial, flat
geometry.  If, as Max Born argued, momentum space is curved, spacetime is just as
observer dependent as space, and the invariant arena for classical physics
is phase space.

So, look around. Do you ``see" spacetime? or do you ``see" phase space?  It is up to
experiment to decide.

}

\newpage

\section*{Acknowledgements}

We are very grateful to Stephon Alexander, Michele Arzano, James
Bjorken, Florian Girelli, Sabine Hossenfelder, Viqar Husain, Etera
Livine, Seth Major, Djorje Minic, Carlo Rovelli, Frederic Schuller
and William Unruh for conversations and encouragement.  GAC and JKG
thank Perimeter Institute for hospitality during their visits in
September 2010, when the main idea of this paper was conceived.
JKG was supported in part by grant 182/N-QGG/2008/0
The work of GAC was supported in part by grant RFP2-08-02
from The Foundational Questions Institute (fqxi.org).   Research at
Perimeter Institute for Theoretical Physics is supported in part by
the Government of Canada through NSERC and by the Province of
Ontario through MRI.

\end{document}